\documentclass[letter,longauth,traditabstract]{aa} 
\usepackage{graphicx}
\usepackage{epstopdf}
\usepackage{natbib}
\usepackage{url}
\begin{document}
\authorrunning{M.~Vaccari et al.}
\titlerunning{The HerMES SPIRE submillimeter local luminosity function}
\title{The HerMES SPIRE submillimeter local luminosity function\thanks{\textit{Herschel} is an ESA space observatory with science instruments provided by European-led Principal Investigator consortia and with important participation from NASA.}}
%
\author{M.~Vaccari\inst{1}
\and L.~Marchetti\inst{1}
\and A.~Franceschini\inst{1}
\and B.~Altieri\inst{2}
\and A.~Amblard\inst{3}
\and V.~Arumugam\inst{4}
\and R.~Auld\inst{5}
\and H.~Aussel\inst{6}
\and T.~Babbedge\inst{7}
\and A.~Blain\inst{8}
\and J.~Bock\inst{8,9}
\and A.~Boselli\inst{10}
\and V.~Buat\inst{10}
\and D.~Burgarella\inst{10}
\and N.~Castro-Rodr{\'\i}guez\inst{11,12}
\and A.~Cava\inst{11,12}
\and P.~Chanial\inst{7}
\and D.L.~Clements\inst{7}
\and A.~Conley\inst{13}
\and L.~Conversi\inst{2}
\and A.~Cooray\inst{3,8}
\and C.D.~Dowell\inst{8,9}
\and E.~Dwek\inst{14}
\and S.~Dye\inst{5}
\and S.~Eales\inst{5}
\and D.~Elbaz\inst{6}
\and D.~Farrah\inst{15}
\and M.~Fox\inst{7}
\and W.~Gear\inst{5}
\and J.~Glenn\inst{13}
\and E.A.~Gonz\'alez~Solares\inst{16}
\and M.~Griffin\inst{5}
\and M.~Halpern\inst{17}
\and E.~Hatziminaoglou\inst{18}
\and J.~Huang\inst{19}
\and E.~Ibar\inst{20}
\and K.~Isaak\inst{5}
\and R.J.~Ivison\inst{20,4}
\and G.~Lagache\inst{21}
\and L.~Levenson\inst{8,9}
\and N.~Lu\inst{8,22}
\and S.~Madden\inst{6}
\and B.~Maffei\inst{23}
\and G.~Mainetti\inst{1}
\and A.M.J.~Mortier\inst{7}
\and H.T.~Nguyen\inst{9,8}
\and B.~O'Halloran\inst{7}
\and S.J.~Oliver\inst{15}
\and A.~Omont\inst{24}
\and M.J.~Page\inst{25}
\and P.~Panuzzo\inst{6}
\and A.~Papageorgiou\inst{5}
\and C.P.~Pearson\inst{26,27}
\and I.~P{\'e}rez-Fournon\inst{11,12}
\and M.~Pohlen\inst{5}
\and J.I.~Rawlings\inst{25}
\and G.~Raymond\inst{5}
\and D.~Rigopoulou\inst{26,28}
\and D.~Rizzo\inst{7}
\and G.~Rodighiero\inst{1}
\and I.G.~Roseboom\inst{15}
\and M.~Rowan-Robinson\inst{7}
\and M.~S\'anchez Portal\inst{2}
\and B.~Schulz\inst{8,22}
\and Douglas~Scott\inst{17}
\and N.~Seymour\inst{25}
\and D.L.~Shupe\inst{8,22}
\and A.J.~Smith\inst{15}
\and J.A.~Stevens\inst{29}
\and M.~Symeonidis\inst{25}
\and M.~Trichas\inst{7}
\and K.E.~Tugwell\inst{25}
\and E.~Valiante\inst{17}
\and I.~Valtchanov\inst{2}
\and L.~Vigroux\inst{24}
\and L.~Wang\inst{15}
\and R.~Ward\inst{15}
\and G.~Wright\inst{20}
\and C.K.~Xu\inst{8,22}
\and M.~Zemcov\inst{8,9}}
\institute{Dipartimento di Astronomia, Universit\`{a} di Padova, vicolo Osservatorio, 3, 35122 Padova, Italy\\
 \email{mattia@mattiavaccari.net}
\and Herschel Science Centre, European Space Astronomy Centre, Villanueva de la Ca\~nada, 28691 Madrid, Spain
\and Dept. of Physics \& Astronomy, University of California, Irvine, CA 92697, USA
\and Institute for Astronomy, University of Edinburgh, Royal Observatory, Blackford Hill, Edinburgh EH9 3HJ, UK
\and Cardiff School of Physics and Astronomy, Cardiff University, Queens Buildings, The Parade, Cardiff CF24 3AA, UK
\and Laboratoire AIM-Paris-Saclay, CEA/DSM/Irfu - CNRS - Universit\'e Paris Diderot, CE-Saclay, pt courrier 131, F-91191 Gif-sur-Yvette, France
\and Astrophysics Group, Imperial College London, Blackett Laboratory, Prince Consort Road, London SW7 2AZ, UK
\and California Institute of Technology, 1200 E. California Blvd., Pasadena, CA 91125, USA
\and Jet Propulsion Laboratory, 4800 Oak Grove Drive, Pasadena, CA 91109, USA
\and Laboratoire d'Astrophysique de Marseille, OAMP, Universit\'e Aix-marseille, CNRS, 38 rue Fr\'ed\'eric Joliot-Curie, 13388 Marseille cedex 13, France
\and Instituto de Astrof{\'\i}sica de Canarias (IAC), E-38200 La Laguna, Tenerife, Spain
\and Departamento de Astrof{\'\i}sica, Universidad de La Laguna (ULL), E-38205 La Laguna, Tenerife, Spain
\and Dept. of Astrophysical and Planetary Sciences, CASA 389-UCB, University of Colorado, Boulder, CO 80309, USA
\and Observational  Cosmology Lab, Code 665, NASA Goddard Space Flight  Center, Greenbelt, MD 20771, USA
\and Astronomy Centre, Dept. of Physics \& Astronomy, University of Sussex, Brighton BN1 9QH, UK
\and Institute of Astronomy, University of Cambridge, Madingley Road, Cambridge CB3 0HA, UK
\and Department of Physics \& Astronomy, University of British Columbia, 6224 Agricultural Road, Vancouver, BC V6T~1Z1, Canada
\and ESO, Karl-Schwarzschild-Str. 2, 85748 Garching bei M\"unchen, Germany
\and Harvard-Smithsonian Center for Astrophysics, MS65, 60 Garden Street,  Cambridge,  MA02138, USA
\and UK Astronomy Technology Centre, Royal Observatory, Blackford Hill, Edinburgh EH9 3HJ, UK
\and Institut d'Astrophysique Spatiale (IAS), b\^atiment 121, Universit\'e Paris-Sud 11 and CNRS (UMR 8617), 91405 Orsay, France
\and Infrared Processing and Analysis Center, MS 100-22, California Institute of Technology, JPL, Pasadena, CA 91125, USA
\and School of Physics and Astronomy, The University of Manchester, Alan Turing Building, Oxford Road, Manchester M13 9PL, UK
\and Institut d'Astrophysique de Paris, UMR 7095, CNRS, UPMC Univ. Paris 06, 98bis boulevard Arago, F-75014 Paris, France
\and Mullard Space Science Laboratory, University College London, Holmbury St. Mary, Dorking, Surrey RH5 6NT, UK
\and Space Science \& Technology Department, Rutherford Appleton Laboratory, Chilton, Didcot, Oxfordshire OX11 0QX, UK
\and Institute for Space Imaging Science, University of Lethbridge, Lethbridge, Alberta, T1K 3M4, Canada
\and Astrophysics, Oxford University, Keble Road, Oxford OX1 3RH, UK
\and Centre for Astrophysics Research, University of Hertfordshire, College Lane, Hatfield, Hertfordshire AL10 9AB, UK}
   \date{Received March 31, 2010; Accepted April 28, 2010}
  \abstract{
  Local luminosity functions are fundamental benchmarks for high-redshift galaxy formation and evolution studies as well as for models describing these processes. Determining the local luminosity function in the submillimeter range can help to better constrain in particular the bolometric luminosity density in the local Universe, and \textit{Herschel} offers the first opportunity to do so in an unbiased way by imaging large sky areas at several submillimeter wavelengths. 
  We present the first \textit{Herschel} measurement of the submillimeter $0<z<0.2$ local luminosity function and infrared bolometric (8-1000 $\mu$m) local luminosity density based on SPIRE data from the HerMES \textit{Herschel} Key Program over 14.7 deg$^2$. 
  Flux measurements in the three SPIRE channels at 250, 350 and 500 $\mu$m are combined with \textit{Spitzer} photometry and archival data. We fit the observed optical-to-submillimeter spectral energy distribution of SPIRE sources and use the $1/V_{max}$ estimator to provide the first constraints on the monochromatic  250, 350  and 500 $\mu$m as well as on the infrared bolometric (8-1000 $\mu$m) local luminosity function based on \textit{Herschel} data. 
  We compare our results with modeling predictions and find a slightly more abundant local submillimeter population than predicted by a number of models. Our measurement of the infrared bolometric (8-1000 $\mu$m) local luminosity function suggests a flat slope at low luminosity, and the inferred local luminosity density, $1.31_{-0.21}^{+0.24}\,\cdot\,10^8\,L_{\odot}\mathrm{Mpc^{-3}}$, is consistent with the range of values reported in recent literature. 
  }
   \keywords{Galaxies: luminosity function -- Galaxies: evolution -- Galaxies: statistics  -- Submillimeter: galaxies}
   \maketitle
\section{Introduction}
Local luminosity functions (LLFs) are essential for the interpretation of the evolution of cosmic sources at any wavelength. While the determination of high-redshift LFs requires deep observations, LLFs can exploit shallower wider-area sky maps. In addition, the extremely careful multi-wavelength identification work required for high-redshift sources is greatly simplified by the lower areal density of bright local sources.
%
The pioneering exploration of the IR sky by the IRAS satellite revealed the rapid evolution of the infrared (IR) LLF (Saunders et al. 1990), illustrating the importance of local studies at infrared (IR) wavelengths. ISO was later able to follow the evolution of the IR LF up to $z\simeq1$ (Pozzi et al. 2004), while \textit{Spitzer} studies using the MIPS 24 $\mu$m (Le Floc'h et al. 2005, Marleau et al. 2007, Rodighiero et al. 2010) and 70 $\mu$m (Frayer et al. 2006, Huynh et al. 2007, Magnelli et al. 2009) channels have both extended our reach to higher redshifts and improved our constraints on the IR bolometric (8-1000 $\mu$m) luminosity of detected sources.
%
%
Due to the moderate volume per unit sky area that one can probe at low redshifts, an essential pre-requisite for determining the LLF is the imaging of large fields. This has so far been difficult for ground-based submillimeter (SMM) cameras, and for example Dunne et al. (2000) carried out SCUBA observations of IRAS bright galaxies instead. It was only recently that BLAST was able to map large areas at SMM wavelengths and thus to first determine the evolution of the SMM LLF (Eales et al. 2009).
The much improved mapping speed and angular resolution of the SPIRE (Griffin et al. 2010) instrument onboard the \textit{Herschel} Space Observatory (Pilbratt et al. 2010) is exploited by the \textit{Herschel} Multi-tiered Extragalactic Survey (HerMES, \url{http://hermes.sussex.ac.uk}) Key Program (Oliver et al. 2010, in prep), which will eventually cover 70 deg$^2$.
While other early \textit{Herschel} studies are probing the evolution of SMM number counts and LFs up to high redshifts (Oliver et al. 2010, Eales et al. 2010), we present a first assessment of the LLF as measured in the three SPIRE bands at 250, 350 and 500 $\mu$m, providing the first robust constraints on the Local Luminosity Density (LLD) at these wavelengths. This study will thus establish a reliable local benchmark at a very early stage of the \textit{Herschel} mission against which models and high-redshift studies can be compared.
Throughout, we adopt a standard cosmology with $\Omega_M=0.28$, $\Omega_\Lambda=0.72$, $H_0=72~\mathrm{km\,s^{-1}\,Mpc^{-1}}$.
\section{Sample selection}
\begin{figure}
   \centering
   \includegraphics[width=0.45\textwidth]{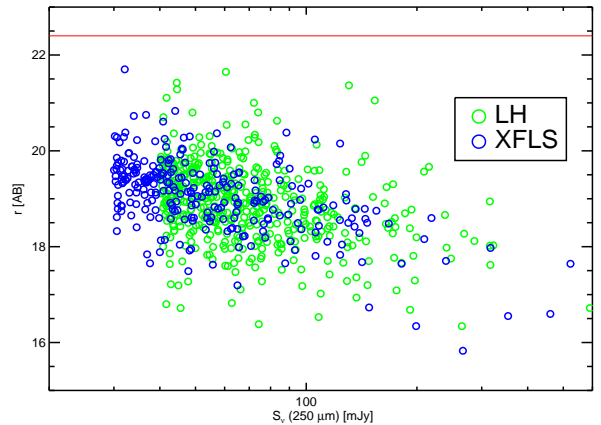} 
   \caption{Optical magnitude vs 250 $\mu$m flux for the full sample. The SDSS $r\simeq22.4$ magnitude limit is indicated by the red line.}
   \label{magr-vs-s250.fig}
\end{figure}

We use a small portion of the observations making up the HerMES Key Program, namely those taken during the \textit{Herschel} Science Demonstration Phase (SDP) in September-October 2009 (Oliver et al. 2010, in prep).
We employ an \textit{Herschel} source catalog obtained following a technique newly developed within the HerMES consortium. To minimize the effect of source blending, the SPIRE fluxes are estimated via linear inversion methods, using the positions of known 24 $\mu$m sources as a prior. The full method and resulting catalogs are described in Roseboom et al. (2010, in prep). The area jointly covered by \textit{Spitzer} and \textit{Herschel} is of 10.7 deg$^2$ within the Lockman Hole (LH) and 4.0 deg$^2$ within the \textit{Spitzer} Extragalactic First Look Survey (XFLS), a total area of 14.7 deg$^2$.
%
%
We select $S/N>5$ reliable SPIRE detections as defined by Roseboom et al. (2010, in prep). While sources not appearing in the input 24 $\mu$m source list will thus not appear in our sample, this incompleteness is negligible at $0<z<0.2$ and the adopted SPIRE flux limits for all but the most extreme SEDs, which we do not expect in significant numbers at low redshift. As discussed by Roseboom et al. 2010 (in prep), the 24 $\mu$m prior thus ensures that the SPIRE catalogs are effectively complete and do not suffer from severe blending issues over the $0<z<0.2$ redshift range down to the adopted flux cuts of 40 mJy and 30 mJy in all bands in LH and XFLS, respectively.
For all Herschel sources, multi-wavelength archival photometry from the ultraviolet to the far-infrared and optical spectroscopy is provided by the \textit{Spitzer}-selected multi-wavelength catalog by Vaccari et al. (2010, in prep) covering the HerMES wide-area fields. 
Available spectroscopic redshifts are mostly from SDSS and NED, but are complemented by those obtained by Huang et al. (2010, in prep) and Sajina et al. (2010, in prep). We use spectroscopic redshifts when available, and photometric redshifts from SDSS DR7 (Abazajian et al. 2009) computed following the neural network technique by Oyaizu et al. (2008). 
Fig.~\ref{magr-vs-s250.fig} shows how optical counterparts to our SMM sources are almost invariably at $r<21$, thus safely within the SDSS nominal magnitude limit of $r \simeq 22.4$.
%
%
%
%
\section{SED fitting}
Because \textit{Herschel} for the first time allows the sampling of the SMM part of the spectrum at multiple wavelengths, one can better constrain the nature of \textit{Herschel} sources by fitting their optical-to-SMM spectral energy distribution (SED) with spectral templates. This allows us not only to determine the required $K$-correction but also to constrain the IR bolometric luminosity of individual sources.
%
For this work, we use SDSS $ugriz$, 2MASS $JHK_S$, IRAC 4-band and MIPS 3-band photometry which are available over the whole 14.7 deg$^2$.
We modeled (see Fig.~\ref{llf-seds.fig}) the 250 $\mu$m sample using the \texttt{hyperz} code of Bolzonella et al. (2000) and the SED templates of Polletta et al. (2007), including the slightly modified versions of a few of these templates introduced by Gruppioni et al. (2010). The latter authors account for the first indications from their SED fitting of \textit{Herschel} sources by introducing {a higher FIR bump which improves the fit to their PACS data points. Keeping the redshift of any given source fixed, we thus determined the $K$-correction at 250, 350 and 500 $\mu$m on the basis of the best-fit SEDs and the IR bolometric (8-1000 $\mu$m) luminosity for the full 250 $\mu$m sample. Interestingly, we find that late-type SED templates such as Sd and Sdm provide the best fit to the majority of sources in our sample. However, a detailed analysis of the LLF as a function of SED type is beyond the scope of this work.
%
%
%
%
\begin{figure}
   \centering
   \includegraphics[width=0.45\textwidth]{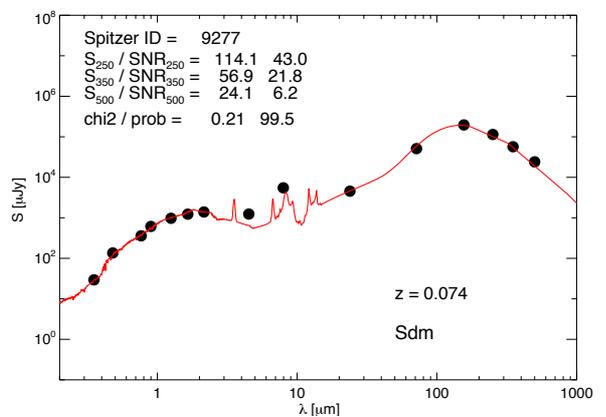}
   \caption{Example of optical-to-SMM SED fitting. The detailed SED fitting allows us to determine the $K$-correction which is then applied on the basis of the best-fit SEDs and the IR bolometric (8-1000 $\mu$m) luminosity for the full 250 $\mu$m sample. We used SDSS, 2MASS, IRAC, MIPS and SPIRE measurements.}
   \label{llf-seds.fig}
\end{figure}
\section{LLF estimation}
In order to determine the monochromatic LLF at 250, 350 and 500 $\mu$m, we employed the standard $1/V_{max}$ LF estimator (Schmidt 1968).
Each source is assigned the maximum redshift $z_{max}$ at which it would still appear within our flux-limited sample and the associated $V_{max}=min\left(V(z<z_{max}) , V(z<0.2)\right)$. For every luminosity bin the LLF estimate is $\phi=\sum_i\frac{1}{V_{max,i}}$ and its associated variance is $\sigma_\phi^2=\sum_i\frac{1}{V_{max,i}^2}$.
Using a similar approach, we then computed the IR Bolometric (8-1000 $\mu$m) LLF integrating the best-fit SED for each source.
For each luminosity bin we then calculated our best LLF estimate and associated error as the weighted average and the associated weighted error of measurements in the two fields.
%
Our estimates of the monochromatic and IR bolometric LLFs are reported in Table~\ref{llf-values.tab}.
We performed a least-squares fit to these measurements with a modified Schechter (or double exponential) function as defined by Saunders et al. (1990)
\begin{equation}
\phi(L) = C\,\left(\frac{L}{L_\star}\right)^{1-\alpha}\,\exp\left[-\frac{1}{2\sigma^2}\,\log^2\left(1+\frac{L}{L_\star}\right)\right]
\label{modschec.eq}
\end{equation}
to determine the IR Bolometric LLD contributed by SPIRE sources through its integration over the full LLF. Best-fit parameters are $\log\,C~\mathrm{[Mpc^{-3}]}=-2.16_{}^{}$,  $\log\,(L_\star/L_\odot)=9.94_{}^{}$, $\alpha=1.00_{}^{}$ and $\sigma=0.50_{}^{}$
and the estimated IR bolometric LLD is $1.31_{-0.21}^{+0.24}\,\cdot\,10^8\,L_{\odot}\mathrm{Mpc^{-3}}$
\section{Results}
%
The LLF predictions from recent literature and previous LLF estimates are compared with our monochromatic and IR bolometric LLF measurements in Fig.~\ref{llf-spire.fig}.
The 250 and 350 $\mu$m LLF measurements in the two fields agree at an average level of 15\,\%, consistent with the levels of cosmic variance predicted by theoretical models for fields of this size at $0<z<0.2$ (Moster et al. 2010). They also agree reasonably well with the BLAST measurements of Eales et al. (2009) but have much greater precision and fewer possible systematic problems. Measurements at 500 $\mu$m, based on a much smaller sample, show a significantly larger variance between the two fields and thus should be treated with caution. Our measurements are slightly above most modeling predictions we considered at all wavelengths, and particularly at the luminosities close to the knee of the LLF which we probed with greater statistical significance. At these luminosities, however, Xu et al. (2001) better reproduce the observed space densities than Negrello et al. (2007) and Franceschini et al. (2010).
%
Our measurements of the IR LLF suggest a flatter low-luminosity slope than previously reported by either Sanders et al. (2003) for an IRAS 60-$\mu$m-selected sample or Rodighiero et al. (2010) for a 24-$\mu$m-selected one, but agrees with predictions by Lagache et al. (2004). However, this flatter slope is mostly driven by our lowest luminosity bin, where the sample size is small and we may start to be affected by incompleteness. Our estimate of $1.31_{-0.21}^{+0.24}\,\cdot\,10^8\,L_{\odot}\mathrm{Mpc^{-3}}$ for the IR LLD at an average redshift of $z\simeq0.1$ excellently agrees with the values obtained by Sanders et al. (2003), Magnelli et al. (2009) and Rodighiero et al. (2010), but is higher than those obtained by Le Floc'h et al. (2005) and Seymour et al. (2010). Previous measurements at both the low and high end of this range in LLD used different combinations of MIPS 24 $\mu$m, IRAS 60 $\mu$m and MIPS 70 $\mu$m selection. While we can now get a better handle on the whole of the SED thanks to using \textit{Herschel} as well as \textit{Spitzer} measurements, the small sample size at both the faint and bright end does not allow us to place stronger constraints on the IR LLD as yet.
%
%
\begin{table}
\centering
\caption{Number of $0 < z < 0.2$ sources used for the determination of LLFs. The number of sources with (spectroscopic/photometric) redshifts is indicated after the total number of sources. The 250 $\mu$m sample is also used to determine the IR bolometric LLF. Field areas are in deg$^2$.}
\begin{tabular}{ccccc}
Field & 250 $\mu$m & 350 $\mu$m & 500 $\mu$m & Area \\
\hline
LH & 478 (369/109) & 116 (99/17) & 18 (16/2) & 10.7 \\
XFLS & 275 (222/53) & 94 (80/14) & 27 (22/5) & 4.0 \\
Total & 753 (591/162) & 210 (179/31) & 45 (38/7) & 14.7 \\
\end{tabular}
\label{llf-numbers.tab}
\end{table}
\begin{table}
\centering
\caption{Monochromatic and IR bolometric LLFs ($0 < z < 0.2$). $L$ indicates $\nu\,L_\nu$ for the monochromatic LLF and $L_{IR}$ for the IR bolometric LLF. $L$ at the bin centers are expressed in $L_\odot$ while LLF estimates and their errors are in $\mathrm{[Mpc^{-3}\,dex^{-1}]}$.}
\begin{tiny}
\begin{tabular}{ccccc}
$\log L$ & $\log\,(\phi,\sigma)_{250}$ & $\log\,(\phi,\sigma)_{350}$ & $\log\,(\phi,\sigma)_{500}$ & $\log\,(\phi,\sigma)_{IR}$ \\
\hline
  7.5  &            -           &           -            & -1.80 ,  -2.69 &           -  \\
  8.0  & -1.97 ,  -2.27 &           -            & -2.12 ,  -2.16 &           -  \\
  8.5  & -2.23 ,  -2.63 & -2.22 ,  -2.92 & -2.84 ,  -2.92 &           -  \\
  9.0  & -2.19 ,  -3.00 & -2.54 ,  -3.05 & -3.36 ,  -3.56 & -2.15 , -3.00 \\
  9.5  & -2.38 ,  -4.50 & -3.14 ,  -3.57 & -4.88 ,  -6.14 & -2.19 , -2.61 \\
 10.0 & -3.14 ,  -4.28 & -4.66 ,  -4.80 &           -            & -2.30 , -2.72 \\
 10.5 & -4.62 ,  -5.23 &           -             &           -           & -2.49 , -3.16 \\
 11.0 &           -            &           -             &           -           & -3.21 , -3.95 \\
 11.5 &           -            &           -             &           -           & -4.24 , -4.65 \\
\end{tabular}
\end{tiny}
\label{llf-values.tab}
\end{table}
%
%
%
\begin{figure*}
   \centering
   \includegraphics[width=0.44\textwidth]{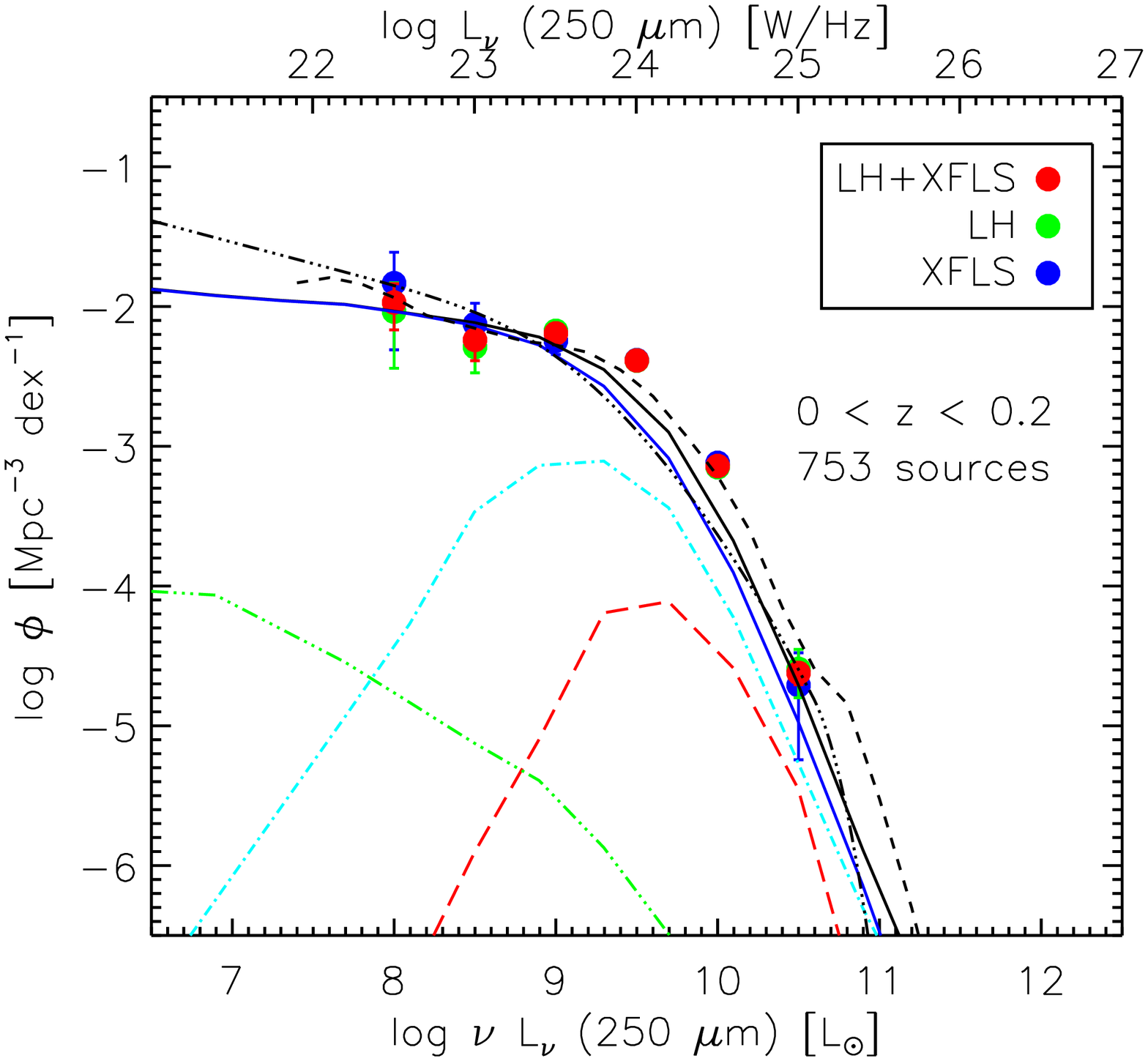}
   \includegraphics[width=0.44\textwidth]{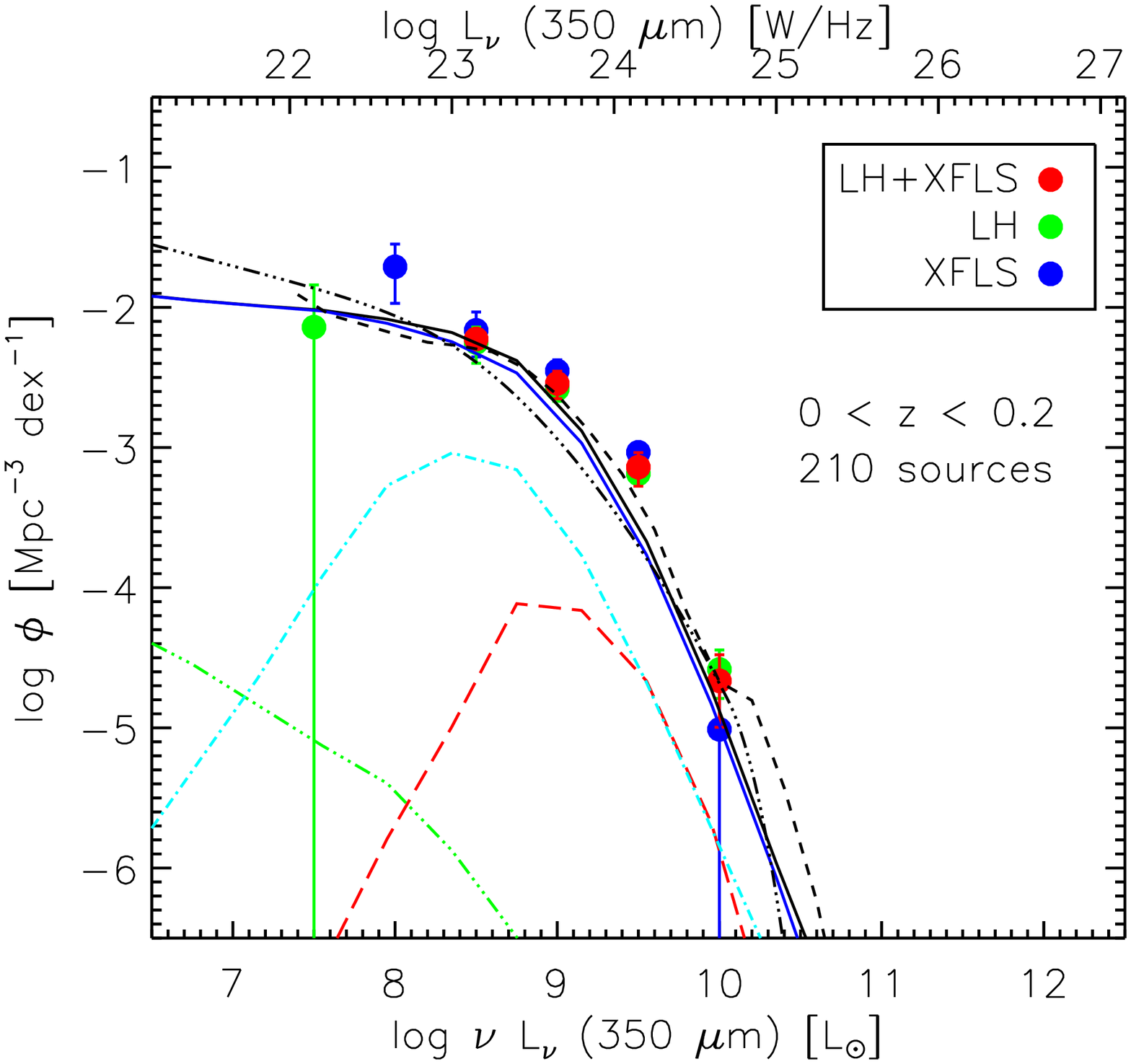}
   \includegraphics[width=0.44\textwidth]{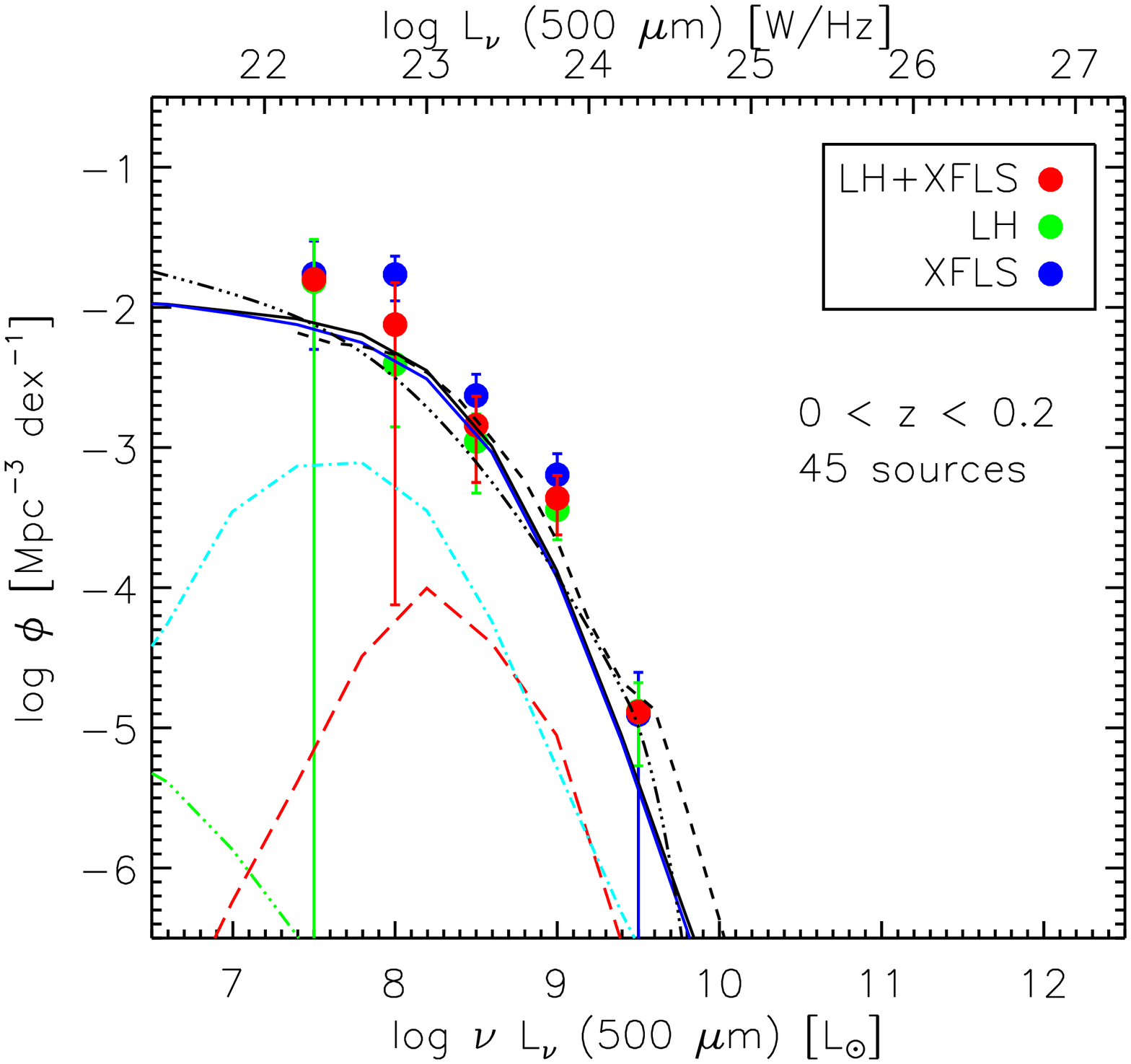}   
   \includegraphics[width=0.44\textwidth]{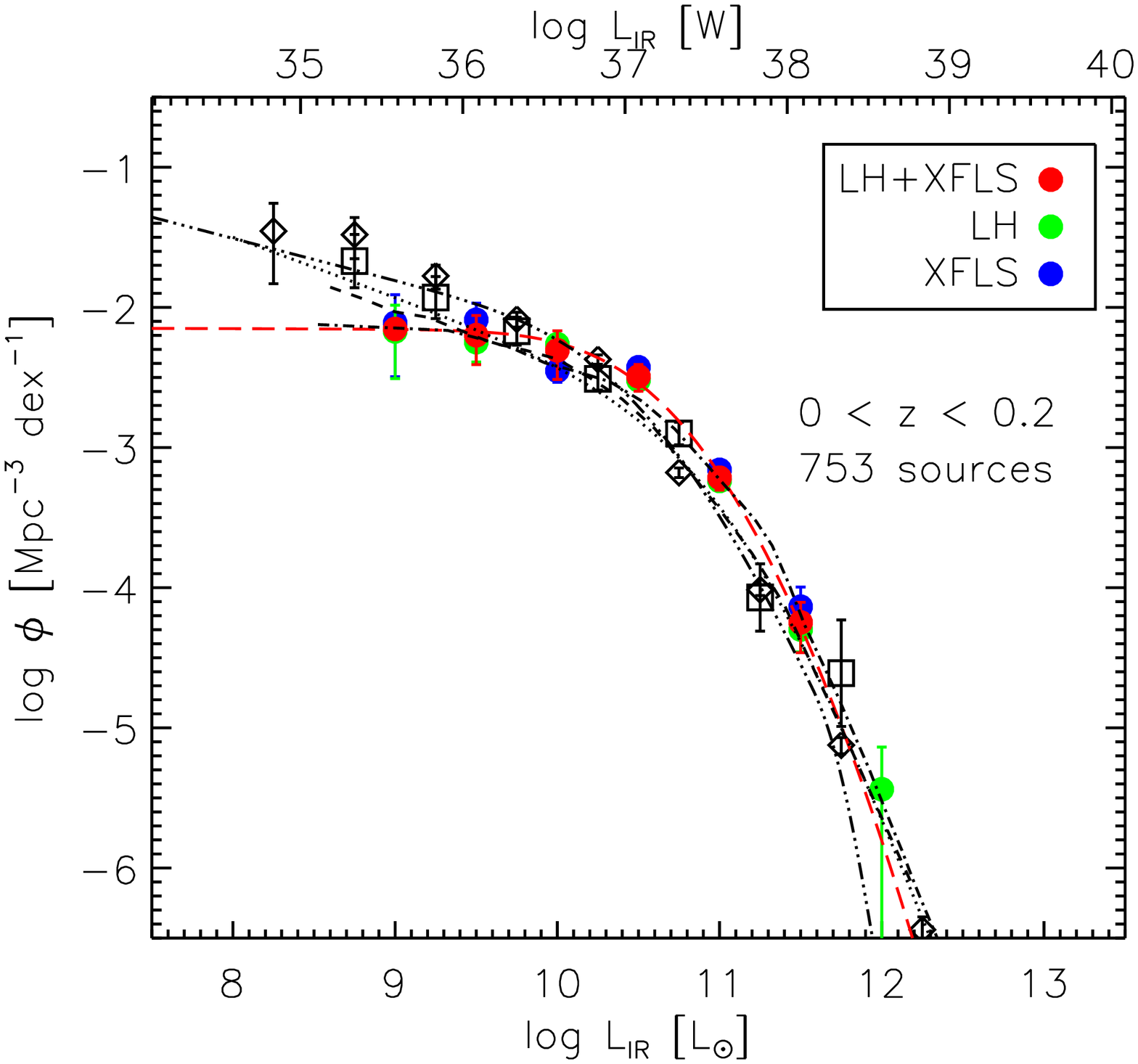}
   \caption{SPIRE 250, 350 and 500 $\mu$m and IR Bolometric LLF in LH+XFLS. In the monochromatic LLF panels, predictions by Xu et al. (2001) and Negrello et al. (2007) are indicated by dashed and dot-dot-dot-dashed lines respectively while model populations by Franceschini et al. (2010) are: dashed blue line are spirals, dot-dashed cyan line are starbursts, dashed red line are high-luminosity starbursts, dot-dot-dot-dashed green line are Type-I AGNs, and the black solid line is the total prediction. Spirals are predicted to make up the bulk of local sources detected by SPIRE, as confirmed by SED fitting. In the IR bolometric LLF panel, the red long-dashed line is the best-fitting modified Schechter (or double exponential) function to the IR bolometric LLF, diamonds and squares are measurements based on the $z\simeq0$ IRAS 60-$\mu$m-selected RBGS by Sanders et al. (2003) and on a $0<z<0.3$ \textit{Spitzer} 24-$\mu$m-selected sample by Rodighiero et al. (2010). Also in the IR Bolometric LLF panel, black dot-dashed, dot-dot-dot-dashed and dotted lines indicate predictions by Lagache et al. (2004), Negrello et al. (2007) and Valiante et al. (2009).
   }
   \label{llf-spire.fig}
\end{figure*}
\section{Conclusions}
Using SPIRE imaging from the HerMES Key Program covering 14.7 deg$^2$ (about 20\% of the total SPIRE area eventually to be covered by HerMES)
we presented the first \textit{Herschel} measurement of the SMM LLF. We computed the IR monochromatic 250, 350 and 500 $\mu$m as well as the IR bolometric (8-1000 $\mu$m) LFs of SPIRE $0<z<0.2$ sources and compared them with modeling predictions for IR LLFs.
Relying on \textit{Spitzer} 24 $\mu$m positions to identify SMM sources and on optical spectroscopic surveys to determine their distances, we placed solid statistical constraints on the SMM LLF and LLD at a very early stage of the \textit{Herschel} mission and compared our results with pre-\textit{Herschel} predictions. We found a slightly more abundant local submillimeter population than predicted by a number of models, a flat slope of the IR LLF at low luminosity and an IR bolometric LLD of $1.31_{-0.21}^{+0.24}\,\cdot\,10^8\,L_{\odot}\mathrm{Mpc^{-3}}$, toward the higher end of values from recent literature. Once HerMES observations are completed, we will be able to better probe the faint and bright end of the SMM LLF and study its evolution with redshift in much greater detail.
%
%
\begin{acknowledgements}
MV acknowledges support from a University of Padova ''\textit{Herschel} \& ALMA'' Fellowship and ASI ''\textit{Herschel} Science'' Contract I/005/07/0.
Micol Bolzonella kindly provided advice on the use of \texttt{hyperz} and Mattia Negrello swiftly produced additional predictions based on his models.
"SPIRE has been developed by a consortium of institutes led by Cardiff University (UK) and including Univ. Lethbridge (Canada); NAOC (China); CEA, LAM (France); IFSI, Univ. Padua (Italy); IAC (Spain); Stockholm Observatory (Sweden); Imperial College London, RAL, UCL-MSSL, UKATC, Univ. Sussex (UK); and Caltech, JPL, NHSC, Univ. Colorado (USA). This development has been supported by national funding agencies: CSA (Canada); NAOC (China); CEA, CNES, CNRS (France); ASI (Italy); MCINN (Spain); Stockholm Observatory (Sweden); STFC (UK); and NASA (USA).
%
%
%
The data presented in this paper will be released through the \textit{Herschel} Database in Marseille (HeDaM, \url{http://hedam.oamp.fr/HerMES}).
\end{acknowledgements}
%

%
\end{document}